\documentclass[aps,pra,superscriptaddress,twocolumn]{revtex4-1}

\usepackage{amsmath,amsfonts,color,graphicx,fullpage, subfig,setspace, floatrow,dcolumn}
\newfloatcommand{capbtabbox}{table}[][\FBwidth]
\usepackage[colorlinks={true},dvipdfm]{hyperref}
\usepackage[singlelinecheck=off,justification=raggedright]{caption}
\hypersetup{citecolor={blue}, filecolor={blue}, linkcolor={blue}, urlcolor={blue}}

\begin{document}
\date{\today}

\author{David Hocker}
\affiliation{Department of Chemistry, Princeton University, Princeton, NJ 08544, USA}
\author{Constantin Brif}
\affiliation{Department of Scalable \& Secure Systems Research, Sandia National Laboratories, Livermore, CA 94550, USA}
\author{Matthew D. Grace}
\affiliation{Department of Scalable \& Secure Systems Research, Sandia National Laboratories, Livermore, CA 94550, USA}
\affiliation{Center for Quantum Information and Control, University of New Mexico, Albuquerque, New Mexico 87131, USA}
\author{Ashley Donovan}
\affiliation{Department of Chemistry, Princeton University, Princeton, NJ 08544, USA}
\author{Tak-San Ho}
\affiliation{Department of Chemistry, Princeton University, Princeton, NJ 08544, USA}
\author{Katharine Moore Tibbetts}
\affiliation{Department of Chemistry, Temple University, Philadelphia, PA 19122, USA}
\affiliation{Department of Chemistry, Princeton University, Princeton, NJ 08544, USA}
\author{Rebing Wu}
\affiliation{Department of Automation, Tsinghua University; Center for Quantum Information Science and Technology, TNList, Beijing 100084, China}
\author{Herschel Rabitz}
\affiliation{Department of Chemistry, Princeton University, Princeton, NJ 08544, USA}

\title{Characterization of control noise effects in optimal  quantum unitary dynamics}
\begin{abstract}
This work develops measures for quantifying the effects of field noise upon targeted unitary transformations. Robustness to noise is assessed in the framework of the quantum control landscape, which is the mapping from the control to the unitary transformation performance measure (quantum gate fidelity). Within that framework, a new geometric interpretation of stochastic noise effects naturally arises, where more robust optimal controls are associated with regions of small overlap between landscape curvature and the noise correlation function. Numerical simulations of this overlap in the context of quantum information processing reveal distinct noise spectral regimes that better support robust control solutions. This perspective shows the dual importance of both noise statistics and the control form for robustness, thereby opening up new avenues of investigation on how to mitigate noise effects  in quantum systems.
\end{abstract}

\maketitle

\section{Introduction}
Controlled quantum systems are being studied for potential applications to many chemical and physical phenomena \cite{brif_jnewphys_2010, Brif.ACP.148.1.2012}. The search for an optimal control can be formulated as an excursion over a control \emph{landscape} specified as the mapping from the controls to a  cost functional (e.g., fidelity). A primary goal is to locate extrema on the landscape that correspond to the best possible fidelity. Under reasonable assumptions about system controllability and dynamical surjectivity, as well as the availability of suitable control resources, the landscape possesses a topology free of suboptimal extrema, enabling a ``trap-free"  search with gradient ascent algorithms \cite{Ho_pra_2009, Rabitz_pra_2005, Hsieh_pra_2008, Hsieh_jcp_2009, PhysRevA.74.012721}.  Key landscape features affecting search efficiency have been considered \cite{Moore_jchemphys_2008, Moore_pra_2011, PhysRevA.84.012109}, and recent work has also examined how constraining critical control resources may hinder the ability to obtain optimal fidelity \cite{moore_pra_2012,donovan_jmatchem_2013, moore_jcp_2012,Riviello_2014}. 

An important issue when considering quantum control is the  extent that noise affects optimal performance. Here we develop a perspective about the influence of random field noise that is based upon the structural features of the landscape.  Optimal control solutions lie at the desired extrema of the control landscape; however, solutions that exhibit a high sensitivity to slight changes in the controls will perform poorly when noise is present. Controls that are inherently insensitive to such changes are termed \emph{robust}.

\begin{figure}[t]
\subfloat{\includegraphics[width=\textwidth, height=\textheight, keepaspectratio]{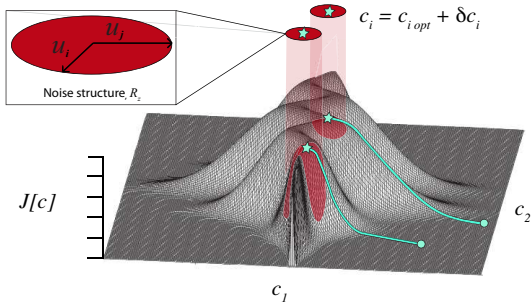}}
\caption{\label{fig:landscape} (color online) Landscape description of robustness. Two control search trajectories ascend to optimal control points, indicated by the stars. Average behavior of the stochastic noise can be specified by noise correlation functions (red ovals), with characteristic directions denoted by the eigenfunctions $u_i$ and $u_j$.  Robustness quality is determined by the degree of landscape curvature that overlaps with eigenfunction directions of the noise correlation function. Better robustness is given by the critical point at the middle of the landscape, which possesses shallow curvature compared to the critical point toward the foreground of the landscape. }
\end{figure}

An optimal control lies in the landscape maximal (or minimal, as appropriate to the objective) critical point submanifold where the slope of the landscape vanishes, while robust optimal controls are additionally located in such regions with low curvature. Figure \ref{fig:landscape} illustrates the qualitative differences between robust and nonrobust solutions with a simplified landscape $J[c]$ that depends upon two control variables $c = [c_1, c_2]$, although practical cases will generally have many variables. Controls are optimized by climbing the landscape from an initial point, indicated by either of the two dots at the base of the landscape, to an optimal solution denoted by a star. In practice, controls that maximize a functional $J$ are inherently subjected to noise that perturbs the controls, reflected in a domain on the landscape indicated by the red oval describing the noise correlation function regions in Fig. \ref{fig:landscape}. Projecting the noise correlation function upon the control landscape indicates the sensitivity of the fidelity to this noise. These robustness concepts apply to any quantum control application, and here we focus on the goal of generating particular unitary transformations. In this regard, an application of special importance is the implementation of gate operations for quantum information processing (QIP). A variety of control methods have been developed to deal with disturbances in QIP such as dynamical decoupling \cite{viola_prl_1999, viola_pra_1998}, dynamically corrected gates \cite{Viola_prl_2009, PhysRevA.86.042329}, and techniques for the correction of systematic noise \cite{ PhysRevA.70.052318, Suzuki1992387}. Instead of focusing on these particular forms of control to assess robustness, we rather seek to investigate features of noise and landscape structure that are encountered for \emph{any} control method.

Assessing robustness in this way is rooted in classical control theory, where higher-order moments of a given fidelity objective are taken as estimates to changes due to noise \cite{robust.control.classical.1990, Nagy2004411, 62265}. Its extension into a quantum context often uses a Magnus expansion of the fidelity objective, where expectation values over noise are taken for higher-order Magnus terms \cite{PhysRevA.90.012316, Green.NJP.9.095004.2013}. Such an approach draws on filter function theory, where components of system dynamics act as a filter upon the noise spectral density. The effects of the magnitude and structure of the noise have also been studied for semiclassical disturbances to the system  \cite{PhysRevLett.87.270405, Zhang_2007, Heule.PRA.82.052333.2010, Heule.EPJD.63.41.2011,Negretti.JPB.44.154012.2011}, as well as for fully quantum mechanical disturbances \cite{PhysRevLett.98.100504}. The relationship between controllability and robustness has also been explored \cite{Khasin.Kosloff.PRL.106.123002.2011, Kallush.njphys.1308.6128.2013}, as have investigations into the dynamical nature of robust control operation \cite{koswara.arXiv.1409.8096.2014}.

Formulating robustness through the lens of the quantum control landscape and its Hessian (see Section \ref{sec:robustness}), as opposed to a Magnus expansion approach as in \cite{PhysRevA.90.012316, Green.NJP.9.095004.2013}, geometrically reveals how robust controls can exist even in the presence of seemingly adverse noise sources for any type of control scheme. The robustness measure given in Section \ref{sec:robustness} is reminiscent of the filter function approaches in refs. \cite{PhysRevA.90.012316, Green.NJP.9.095004.2013}; however, the use of the landscape Hessian (which is highly nonlinear in the controls) naturally reflects the system dynamics and acts as a filter that directly reveals the subtle noise-system relationship.  While the landscape Hessian's role in robustness was previously identified \cite{Ho_pra_2009,Moore_pra_2011}, the general implications of its relationship with noise structure have not yet been addressed. 

This paper quantitatively investigates the spectral relationship between the Hessian and the noise in a general manner, revealing specific spectral regimes of noise that can either hinder or support robust controls. Doing so provides a foundation for optimization studies, including Pareto tradeoffs, and further examination of the role of quantum control landscape features in this regard. The structure of the paper is organized as follows: Section \ref{sec:formulation} outlines the formalism of optimal unitary transformation control. Section \ref{sec:robustness} develops the formalism of a Hessian-based robustness measure. Analytical and numerical features of robustness for different noise types are examined in section \ref{sec:numerical}, followed by concluding remarks in section \ref{sec:conclusion}.

\section{Unitary control objective}
\label{sec:formulation}
Consider an $N$-level quantum system with a Hamiltonian expressed  as 
\begin{align}
\label{eq:hamil}
H(t) = H_0 + \mu\epsilon(t),
\end{align}
where $H_0$ is the field-free Hamiltonian, $\mu$ is the dipole, and $\epsilon(t)$ is a control field. The Hamiltonian generates a unitary propagator $U(t) \equiv U(t,0)$ satisfying the Schr\"odinger equation:
\begin{align}
\label{eq:tdse}
i\frac{\partial}{\partial t} U(t) &= H(t) U(t),
\end{align}
 where  $U(0) = \mathbb{I}$, and $\hbar = 1$. The solution $U(t)$ may be written as
\begin{align}
\label{eq:timeorderedU}
U(t) &= \hat{\mathcal{T}}\text{exp}\left[ -i\int_0^t H(t') dt' \right], 
\end{align} 
where $\hat{\mathcal{T}}$ is the time-ordering operator \cite{Schwabl_monograph}. 

The performance of the final controlled transformation for performing the target unitary gate $W$, at time $T$, can be quantified by the cost functional $J$ that depends upon the control field $\epsilon(t)$ \cite{Ho_pra_2009}:
\begin{align}
\label{eq:J}
J[\epsilon(t)] &= \frac{1}{4N}\|W-U(T)\|^2 \nonumber \\
&= \frac{1}{2}-\frac{1}{2N}\text{Re}(\text{Tr}[W^{\dag}U(T)]),
\end{align}
where $\|A\| \equiv \sqrt{ \text{Tr} \left [ A^{\dag}A \right ]}$ is the Hilbert-Schmidt norm for a matrix $A$. The fidelity of a performed transformation is then taken as $\mathcal{F} = 1-J$. For this objective, the goal is to minimize $J$ (rather than maximize as shown in Figure \ref{fig:landscape}) such that an optimal control generates $U(T) = W$ ($J= 0$), while a worst-case control generates $U(T)= -W$ ($J_{\max}=1$). Unitary transformations that differ only by a global phase are physically indistinguishable as gate operations, and a phase-independent version of the functional in Eq.\ref{eq:J} can be used \cite{PhysRevA.68.062308}. As the landscape features are predominantly developed for the functional in Eq. (\ref{eq:J}), and the phase of the target transformation can bear significant implications for time-optimal control strategies \cite{moore_pra_2012}, we focus here on the phase-dependent form.

Locating an optimal control for $J$ through gradient-based methods involves descending the control landscape to find minimal critical points of $J$, where the gradient of $J$ with respect to the control is zero. There are $N+1$ critical submanifolds at equally spaced values of $J = 0,1/N,2/N,...N$ \cite{Ho_pra_2009}. Under the assumptions that i) the system is controllable,  ii)  the time-dependent coupling matrix $\mu(t) = U^{\dag}(t) \mu U(t)$ is full rank, and  iii) no constraints are placed upon the controls, the landscape possesses a favorable trap-free topology, and contains only a global maximum and minimum, with the other critical points of $J$ corresponding to saddles \cite{Ho_pra_2009, Rabitz_science_2004}. The overwhelming numerical and experimental evidence suggests that these assumptions are generally satisfied, at least to a practical level, for physically applicable control schemes \cite{Rabitz_prl_2012,Moore_pra_2011, Pechen_prl_2011}.

The gradient of $J$ in Eq.(\ref{eq:J}) is
\begin{align}
\label{eq:delJdelE_1}
\frac{\delta J}{\delta \epsilon(t)} &= -\frac{1}{2N}\text{Re}\left( \text{Tr}\left[ W^{\dag}\frac{\delta U(T)}{\delta \epsilon(t)}\right ] \right ),
\end{align}
and utilizing Eq. (\ref{eq:hamil}) we have \cite{Ho_jphotochem_2006}
\begin{align}
\label{eq:grad}
\frac{\delta J}{\delta \epsilon(t)} &= -\frac{1}{2N}\text{Im}(\text{Tr}[W^{\dag}U(T)\mu(t)]).
\end{align}
A critical point is characterized by its Hessian,
\begin{align}
\label{eq:hess_1field}
\mathcal{H}_{\epsilon}(t,t') &= \frac{\delta}{\delta \epsilon(t')}\left[ \frac{\delta J}{\delta \epsilon(t) }\right ] = \frac{\delta^2 J}{\delta \epsilon(t') \delta \epsilon(t)} \nonumber \\
&=\frac{1}{2N}\text{Re}(\text{Tr}[W^{\dag}U(T)\mu(t')\mu(t)]) \quad t'  \geq t,
\end{align}which specifies the landscape curvature \footnote{The Hessian expression in Eq. (\ref{eq:hess_1field}) is valid anywhere on the landscape, even away from critical points.}.
The Hessian may be expressed in an eigen-decomposition,
\begin{align}
\label{eq:hess_eig}
\mathcal{H}_{\epsilon}(t,t') &= \sum_{i=1}^{N^2-1} \lambda_{i} v_{i}(t) v_{i}(t'),
\end{align}
where the eigenfunctions $\{v_{i}(t)\}$ give the principle directions of curvature and the non-zero eigenvalues $\{\lambda_{i}\}$ weight their contributions \cite{Ho_pra_2009,Moore_pra_2011}. The Hessian also has an accompanying infinite dimensional nullspace, which is important for understanding the influence of noise upon optimality \cite{Rabitz_pra_2005, Ho_pra_2009, Hsieh_pra_2008}.  Note that the Hessian at any point on the landscape is bounded by
\begin{align}
\label{eq:Hessian_bounds}
\left | \mathcal{H}_{\epsilon}(t,t') \right | \leq \frac{1}{2N} \|\mu \|^2.
\end{align}
The trace of the Hessian at points of optimality  is invariant to the control,  
\begin{align}
\label{eq:Hess_fixed}
Tr[\mathcal{H}_{\epsilon}] &= \int_0^T \mathcal{H}_{\epsilon}(t,t) dt \nonumber \\
&= \frac{1}{2N} T \text{Re} \left ( \text{Tr} \left [ \mu^2 \right ] \right ) \nonumber \\
&= \frac{1}{2N} T \|\mu \| ^2.
\end{align}
This property has important implications for seeking robust controls, as the overall magnitude of the Hessian may not be reduced. This situation is in stark contrast to the very favorable circumstances for the control of state-to-state transformations, where the magnitude of the Hessian can be freely manipulated by appropriate variation of the control field \cite{Beltrani_jphysb_2011}.

\section{Formulation of the robustness measure}
\label{sec:robustness}
Robustness to noise about an optimal control $\epsilon(t)$ is dominated by the second-order contribution to the Taylor series expansion of $J$, assuming that the perturbation $\delta \epsilon(t)$ is small (i.e., in the weak noise approximation):
\begin{align}
\label{eq:taylor_series}
\delta ^2 J[\epsilon(t) + \delta \epsilon(t)]  &= \frac{1}{2} \int_0^T\int_0^T \mathcal{H}_{\epsilon}(t,t') \delta \epsilon(t) \delta \epsilon(t') dt dt'.
\end{align}
Noise in the control can be expressed as entering in either of two distinct forms: additively ($\epsilon(t)  \rightarrow \epsilon(t) + \delta \epsilon(t)$) or multiplicatively  ($\epsilon(t)  \rightarrow \epsilon(t) [1 + \delta \epsilon(t)]$). In practice the noise may have both contributions present. For convenience we will separately consider these two forms. Since the noise arises due to a stochastic process, it is appropriate to take the statistical expectation value of Eq. (\ref{eq:taylor_series}) with respect to the probability distribution of the corresponding noise process to give a measure of robustness:
\begin{align}
\label{eq:Kba}
K_{A} &= E\{\delta^2 J_{A}\} \nonumber \\
 &= \frac{1}{2} \int_0^T \int_0^T \mathcal{H}_\epsilon(t,t') R(t,t') dt dt', \\
\label{eq:Kbm}
K_{M} &= E\{\delta^2 J_{M}\} \nonumber \\
&= \frac{1}{2} \int_0^T \int_0^T \mathcal{H}_\epsilon(t,t') \epsilon(t)\epsilon(t') R(t,t') dt dt'.
\end{align}
Here, $R(t,t') = E\{\delta \epsilon(t)\delta \epsilon(t')\}$ is the noise correlation function of $\delta \epsilon(t)$, and the subscripts on $K$ denote additive ($A$) or multiplicative ($M$) noise. Good robustness of a control is indicated by $K_{A}$ or $K_{M}$ being small.

For wide-sense stationary (WSS) noise processes, where the mean and standard deviation of the probability distribution characterizing the noise are constant in time, a complimentary view of robustness can be presented conveniently in the frequency domain \cite{Brockwell_monograph}. The noise correlation function $R(t,t') = R(\tau)$ of a WSS process only depends upon the time difference $\tau \equiv t-t'$. The Wiener-Khinchin theorem relates the noise correlation function of a WSS noise signal to its power spectral density  $S(\omega)$, which yields
\begin{align}
\label{eq:constantin_Fdomain_add}
K_{A} &= \frac{1}{4\pi} \int_{-\infty}^{\infty} \mathcal{H}_{\epsilon}(\omega) S(\omega) d\omega, 
\end{align}
where
\begin{align}
\label{eq:constantin_Fdomain_add1}
\mathcal{H}_{\epsilon}(\omega)&= \int_0^T \int_0^T \mathcal{H}_{\epsilon}(t,t') e^{ i \omega (t-t')} dt' dt, \\
S(\omega) &= \int_{-\infty}^{\infty} R(\tau) e^{- i \omega \tau} d\tau,
\end{align}
and for multiplicative noise
\begin{align}
\label{eq:constantin_Fdomain_mult}
K_{M} &= \frac{1}{4\pi} \int_{-\infty}^{\infty} \tilde{\mathcal{H}}_{\epsilon}(\omega) S(\omega) d\omega,
\end{align}
where
\begin{align}
\label{eq:constantin_Fdomain_mult1}
\tilde{\mathcal{H}}_{\epsilon}(\omega)&= \int_0^T \int_0^T \mathcal{H}_{\epsilon}(t,t') \epsilon(t)\epsilon(t') e^{ i \omega (t-t')} dt' dt .
\end{align}
The landscape interpretation of robustness presented in Fig. \ref{fig:landscape} can be understood in terms of an eigen-decomposition of both the Hessian in Eq. (\ref{eq:hess_eig}) and the noise correlation function as 
\begin{align}
\label{eq:AF_decomp}
R(t,t') &= \sum_{j=1}^{M} \gamma_j  u_j(t')u_j(t),
\end{align}
where $\gamma_j$ are the eigenvalues, $u_j(t)$ are the eigenfunctions, and $M$ is the rank of $R$ (either finite or infinite depending upon the specific noise process). The eigenfunctions $u_j(t)$ and $v_i(t)$ are taken as real. Combining Eq. (\ref{eq:AF_decomp}) with the Hessian expression in Eq. (\ref{eq:hess_eig}), the robustness measures can be written in terms of a set of overlap coefficients  $C_{A_{i, j}}$ and $C_{M_{i, j}}$, for additive and multiplicative noise, respectively: 
\begin{align}
\label{eq:K_eigendecomp_general1}
K_{A} & = \frac{1}{2} \sum_{i=1}^{N^2-1}  \sum_{j=1}^{M}\lambda_{i} \gamma_j \left (\int_0^T v_i(t)u_j(t)dt \right)^2 \nonumber \\
&= \frac{1}{2} \sum_{i=1}^{N^2-1}  \sum_{j=1}^{M}\lambda_{i} \gamma_j C_{A_{i, j}}, 
 \end{align}
\begin{align}
\label{eq:K_eigendecomp_general2}
K_{M} &  = \frac{1}{2} \sum_{i=1}^{N^2-1}  \sum_{j=1}^{M} \lambda_{i} \gamma_j \left (\int_0^T v_i(t)u_j(t)\epsilon(t)dt \right)^2 \nonumber \\
&= \frac{1}{2} \sum_{i=1}^{N^2-1}  \sum_{j=1}^{M} \lambda_{i} \gamma_j C_{M_{i, j}}.
\end{align}When the noise correlation function and Hessian strongly overlap, this can lead to poor robustness. Figure \ref{fig:landscape} visualizes how the overlap contributes to robustness quality, where ovals in that figure represent $R$. Even with significant overlap in the $C$ coefficients in Eqs. (\ref{eq:K_eigendecomp_general1}) and (\ref{eq:K_eigendecomp_general2}), robust controls may still exist in regions where the product of eigenvalues  $\lambda_{i}$ and $\gamma_{j}$ are relatively small.

 Engineering methods to cope with noise often focus on how to compensate for a given noise source tied to the nature of the control in a particular physical system. However, the robustness measures in Eqs. (\ref{eq:Kba}) and  (\ref{eq:Kbm}) or equivalently Eqs. (\ref{eq:constantin_Fdomain_add}) and (\ref{eq:constantin_Fdomain_mult}) emphasize that the character of the engineered system, as well as the features of the noise structure, enter on equal footing. If an optimal control acting on a particular physical system produces a Hessian that overlaps significantly with the associated noise form, then modifying the physical system realization may permit operation under favorable, alternative noise contributions. This route may just as readily lead to better robustness as could an elaborate control scheme that modifies the Hessian structure in the original system. Either approach can successfully enhance robustness, thereby making clear that quantum engineering can beneficially operate with dual consideration of system realization along with operational noise characteristics.
 
 The fundamental relationship between the landscape and noise structures also highlights the complexity of robustness, as accessible directions and associated curvature on the landscape are both coupled to one another through the conserved trace of the Hessian, as well as coupled externally to multiple components of a noise correlation function. The following section numerically examines these intricate relationships between noise and landscape structures, and explores dynamic trends about robustness.

\section{Illustrations of robustness behavior}
\label{sec:numerical}
Certain noise processes may be either naturally difficult or, alternatively, easy to tolerate, and we will discuss these cases to consider the possible factors that influence robustness.  In order to provide a broad assessment, we consider models of one-qubit and two-qubit systems. The first case is a generalized spin-1/2 system with the Hamiltonian
\begin{align}
\label{eq:1qubit_H}
H(t) = \frac{\omega_1}{2} \sigma_z + \frac{\epsilon(t)}{2}\sigma_x,
\end{align}
where $\omega_1$ is the energy level spacing between the $|0\rangle$ and $|1\rangle$ states, and $\sigma_z$ and $\sigma_x$ are Pauli operators. For the two-qubit case, an additional isotropic Heisenberg coupling term between the two qubits is included, along with a separate control field for each qubit,
\begin{align}
\label{eq:2qubit_H}
H(t) &= H_0 + H_1(t) + H_2(t),\\
\label{eq:2qubit_H2}
H_0 &= \frac{\omega_1}{2} \sigma_z^{(1)}+ \frac{\omega_2}{2}\sigma_z^{(2)}  + J_{1,2}\mathbf{\sigma}^{(1)}\cdot \mathbf{\sigma}^{(2)}, \\
\label{eq:2qubit_H3}
H_i(t) &= \frac{\epsilon_i(t)}{2} \sigma_x^{(i)} \quad i = 1,2.
\end{align}
The operators $\mathbf{\sigma}^{(i)} = [\sigma_x^{(i)}, \sigma_y^{(i)}, \sigma_z^{(i)}]$ are tensor products of the one-qubit Pauli matrices with the $2\times2$ identity matrix $ \mathbb{I}_2$:
\begin{align}
\label{eq:pauli_tensor}
\sigma_a^{(1)}  &= \sigma_a \otimes \mathbb{I}_2, & \sigma_a^{(2)} &= \mathbb{I}_2 \otimes \sigma_a, \quad a=x,y,z.
\end{align}
Energy level spacings of $\omega_1 = 20$ and $\omega_2 = 24$ are used with a weak interqubit coupling strength $J_{1,2} = 0.2$. One-qubit operations are conducted over a time interval $T=1$, two-qubit operations over $T=10$, and a temporal resolution $\Delta t=0.01$ is used in solving the Schr\"odinger equation. Propagation is performed through short time steps,
\begin{align}
\label{eq:prop}
U(t+\Delta t) &= e^{ -i H(t+\Delta t)  \Delta t}  U(t).
\end{align}

We consider a decaying exponential noise correlation function corresponding to a $\sim$$1/\omega^2$ noise spectral density,
\begin{align}
\label{eq:acf_forms}
R_z(t,t') &= A^{2}e^{-\frac{|t-t'|}{\alpha}}, & S(\omega) &= \frac{ 2A^2}{\frac{\pi}{\alpha}+\pi\alpha\omega^2},
\end{align}
with a correlation time $\alpha$ characterizing the low-frequency regime ($\alpha \gg 1$) and a white noise-like regime ($\alpha \ll 1$). The noise strength is chosen as  $A^{2}=10^{-4}$. A constant value of $A^2$ is chosen across all values of $\alpha$ to assess the effect of disturbances to controls with an average magnitude of $A = 0.01$ (i.e., $\sim$0.1\% of optimal control field amplitudes in this study). The Hessian in the single qubit case is given by

\begin{align}
\label{eq:robustness_1hess}
 \mathcal{H}_{\epsilon}(t,t') &= \frac{1}{2N}\text{Re}\left ( \text{Tr} \left [ W^{\dag}U(T) \sigma_x(t') \sigma_x(t) \right ] \right ),\\
 \quad t' \geq t .\nonumber
\end{align}
The Hessian for the  two-qubit system is formed in an analogous fashion. We assume that the two fields have independent noise contributions with each expressed by the same noise correlation function. Thus, the total robustness is the contribution from both qubits
\footnote{Correlated noise where a single noise source affects multiple Hamiltonian parameters is possible, and would require consideration of Hessians with nonzero cross terms such as $\delta^2 J/ \delta c_1(t) \delta c_2(t')$.}:
\begin{align}
\label{eq:K-2qubit}
K_{A} =& \frac{1}{2} \int_0^T \int_0^T \left [ \mathcal{H}_{\epsilon_1}(t,t')+ \mathcal{H}_{\epsilon_2}(t,t') \right ] R(t,t') dt dt',
\end{align}
and similarly for $K_M$.

The gate transformations are the one-qubit Hadamard and two-qubit CNOT gate
\begin{align}
\label{eq:cnot_hmard}
W_{H} &= \frac{e^{i \pi/2}}{\sqrt{2}}\left ( \begin{array}{c c}
1&1\\
1&-1\\ \end{array} \right), \\
\label{eq:cnot_hmard2}
 W_{\mathrm{CNOT}} &= e^{i \pi/4}\left ( \begin{array}{c c c c}
1&0&0&0\\
0&1&0&0\\
0&0&0&1\\
0&0&1&0\\ \end{array} \right ). 
\end{align}
A global phase is included in the gate definition in order to ensure that the target transformation is in the special unitary group SU($N$), a requirement for  successful optimization of $J$  given the Hamiltonian structure of Eqs. (\ref{eq:1qubit_H})-(\ref{eq:2qubit_H3}) \cite{Moore_pra_2011}. 

The optimal controls in this study are located through minimization of the distance measure in Eq. (\ref{eq:J}) with the D-MORPH algorithm \cite{Moore_pra_2011,Rothman_pra_2006, Rothman_pra_2005}. The controls depend on the search variable $s \geq 0$ with the requirement that $dJ/ds \leq 0$,
\begin{align}
\label{eq:dmorph1}
\frac{d J}{d s} &= \int_0^T \frac{\delta J}{\delta \epsilon(s,t)}\frac{\partial \epsilon(s,t)}{\partial s} dt \quad \leq 0,
\end{align}
assured by 
\begin{align}
\label{eq:dmorph2}
\frac{\partial \epsilon(s,t)}{\partial s} &= -\frac{\delta J}{\delta \epsilon(s,t)}.
\end{align}
Equation (\ref{eq:dmorph2}) is numerically solved with a fourth-order Runga-Kutta integrator (MATLAB's \texttt{ode45} routine).  For QIP applications, gate fidelity demands are high, and an optimal control is required to create a baseline, optimal $J$ values of $J _0<  10^{-6} $. This degree of optimality is in the regime where error-correcting codes should be operational \cite{Suchara_qec2013}.

\subsection{Distributions of Robustness for Optimal Controls}
\label{eq:distributions}
\begin{figure}[htb]
\begin{center}
\subfloat{\includegraphics[width=\textwidth, height=\textheight, keepaspectratio]{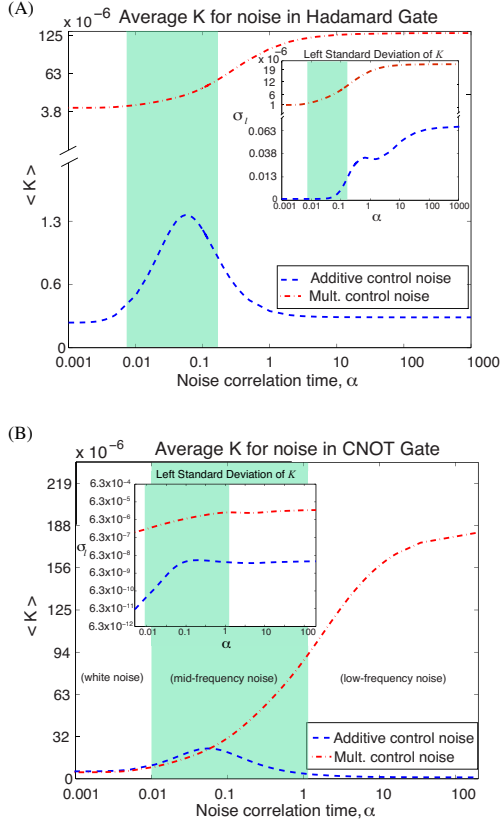}}
\caption{\label{fig:averages} (color online) Averages and left standard deviations (inset) of robustness distributions of the optimal Hadamard gate (A) and CNOT gate (B). The shaded regions denote mid-frequency noise, where noise power is centered around the same spectral domain as system dynamics ($\omega \in [6,50]$). White noise and low-frequency noise lie to the left and right of this region, respectively. }
\end{center}
\end{figure}

A general picture of robustness to noise, as well as any inherent difficulties in improving it, can be gained by examining an ensemble of optimal fields for their robustness quality. To build this ensemble, 1000 initial fields with random amplitudes and phases for on-resonance frequency components were optimized to minimal critical points on the landscape, followed by calculation of robustness measures. After initial choice of the random field, its form is dictated by the optimal solution to Eq. (\ref{eq:dmorph2}). The averages $\langle K_{A/M} \rangle $ and left standard deviations $\sigma_{l}$ of the ensemble for both additive and multiplicative noise types are shown in Figure \ref{fig:averages} (A) for the Hadamard gate, and Figure \ref{fig:averages} (B) for the CNOT gate. Assuming that the ensemble adequately represents the range of possible robustness values for critical points on the landscape, a large left standard deviation indicates a high potential for optimization of robustness against a given noise form. For each of the $L$ optimal fields in the ensemble generating a robustness measure $K_i \leq \langle K \rangle$, $\sigma_l$ is expressed as
\begin{align}
\label{eq:std_l}
\sigma_l &= \left (\frac{1}{L} \sum_i^{L} \left (K_i-\langle K \rangle \right )^2 \right )^{\frac{1}{2}}, \quad K_i \leq \langle K \rangle.
\end{align}

Here, $\langle K \rangle$ is the average change in $J$ due to noise where $\langle J \rangle = J_0 + \langle K \rangle$. When $\langle K \rangle$ $\gtrsim J_0$ and $\sigma_l \ll \langle K \rangle $, then this situation indicates operation under non-robust conditions that may be difficult to improve upon by searching for the most robust control. Such an instance is evident in the case of additive noise in the CNOT gate, where $\langle K \rangle$ is over an order of magnitude larger than $J_0$ for $\alpha \in [0.01,1]$ (shaded region), and $\sigma_l$ was several orders of magnitude smaller than $\langle K \rangle$. However, the Hadamard gate did operate in a nearly robust manner in the presence of additive noise, with $\langle K\rangle$ reaching its maximum of $\sim 10^{-6}$ for $\alpha = 0.06$.   $\langle K \rangle$ and $\sigma_l$ for multiplicative noise are practically invariant to the dimension of the target gate in the present cases.

 A range of correlation time values for which noise power overlaps with the spectrum of system dynamics is shaded in Figs. \ref{fig:averages}(A) and \ref{fig:averages}(B), ranging from $\alpha \in [0.008,0.12]$ for the Hadamard gate, and  $\alpha \in [0.008,1]$ for the CNOT gate. This spectral region where system dynamics are important was identified by examining dominant frequency components in the power spectra of the optimal controls, which displayed significant power in $\omega \in [6,50]$. The corresponding range of $\alpha$ values was characterized by identifying the onset of the low-frequency regime (i.e., more then 90\% noise power density in $\omega < 6$) as well as the white noise regime (noise power density in the window $\omega \in [6,50]$ approaching a constant value). The shaded regimes in Fig. \ref{fig:averages} are referred to as ``mid-frequency" noise, as they lie in between low frequency and white noise.  $\langle K \rangle$ has a maximum in the mid-frequency regime, while for multiplicative noise $\langle K \rangle$ increases monotonically with $\alpha$. $\sigma_l$ decreases dramatically as $\alpha$ decreases, displaying the difficulty for robustness to be improved at small $\alpha$. The standard deviations are similar in magnitude for both gates.

The trends seen in robustness distributions can be further qualified by examining the different spectral regimes of noise where robustness quality is distinct. Comparing the averages and standard deviations for these different regions of $\alpha$ in Fig. \ref{fig:averages}, the mid-frequency regime possesses the most diversity in average robustness, as well as standard deviation within the distributions. This frames mid-frequency noise as having a complex relationship with system dynamics that can be either tolerable or detrimental for performing quantum operations. These three different spectral regimes are further examined for their landscape features in the following sections.

\subsection{Mid-frequency noise}
\label{sec:general_features}
\begin{figure}[htb]
\begin{center}
\subfloat{\includegraphics[width=\textwidth, height=\textheight, keepaspectratio]{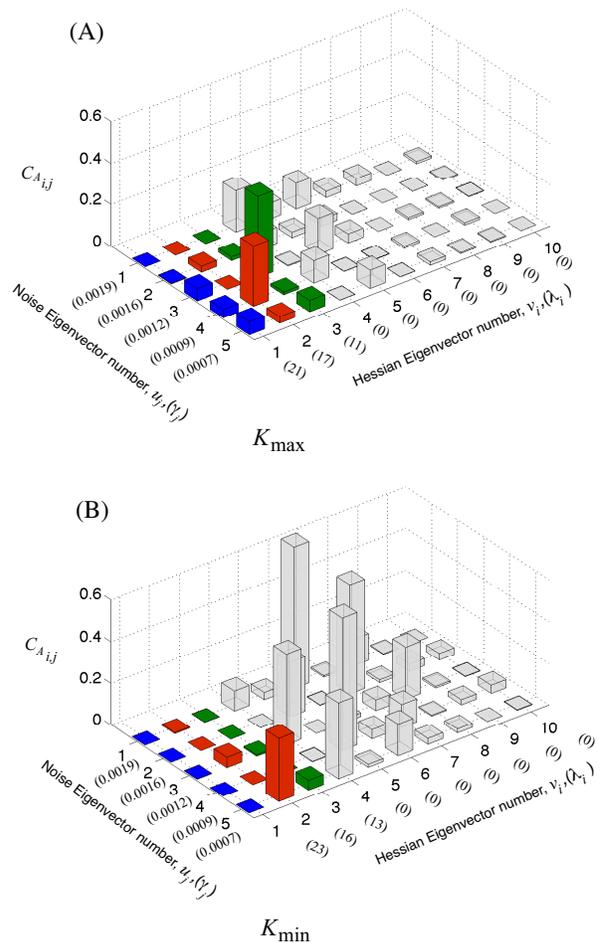}}
\caption{\label{fig:colored_add} (color online) Additive noise and Hessian overlap terms  $C_{A_{i,j}}$ for mid-frequency noise in the Hadamard gate for poor robustness (A) and best robustness (B). The noise correlation time for the noise was $\alpha=0.1$. Listed in parentheses are the associated eigenvalues of each eigenfunction. Robustness is achieved by altering the dynamics such that the noise-Hessian overlap shifts to the Hessian nullspace.}
\end{center}
\end{figure}

\begin{figure*}[htb]
\begin{center}
\subfloat{\includegraphics[width=\textwidth, height=\textheight, keepaspectratio]{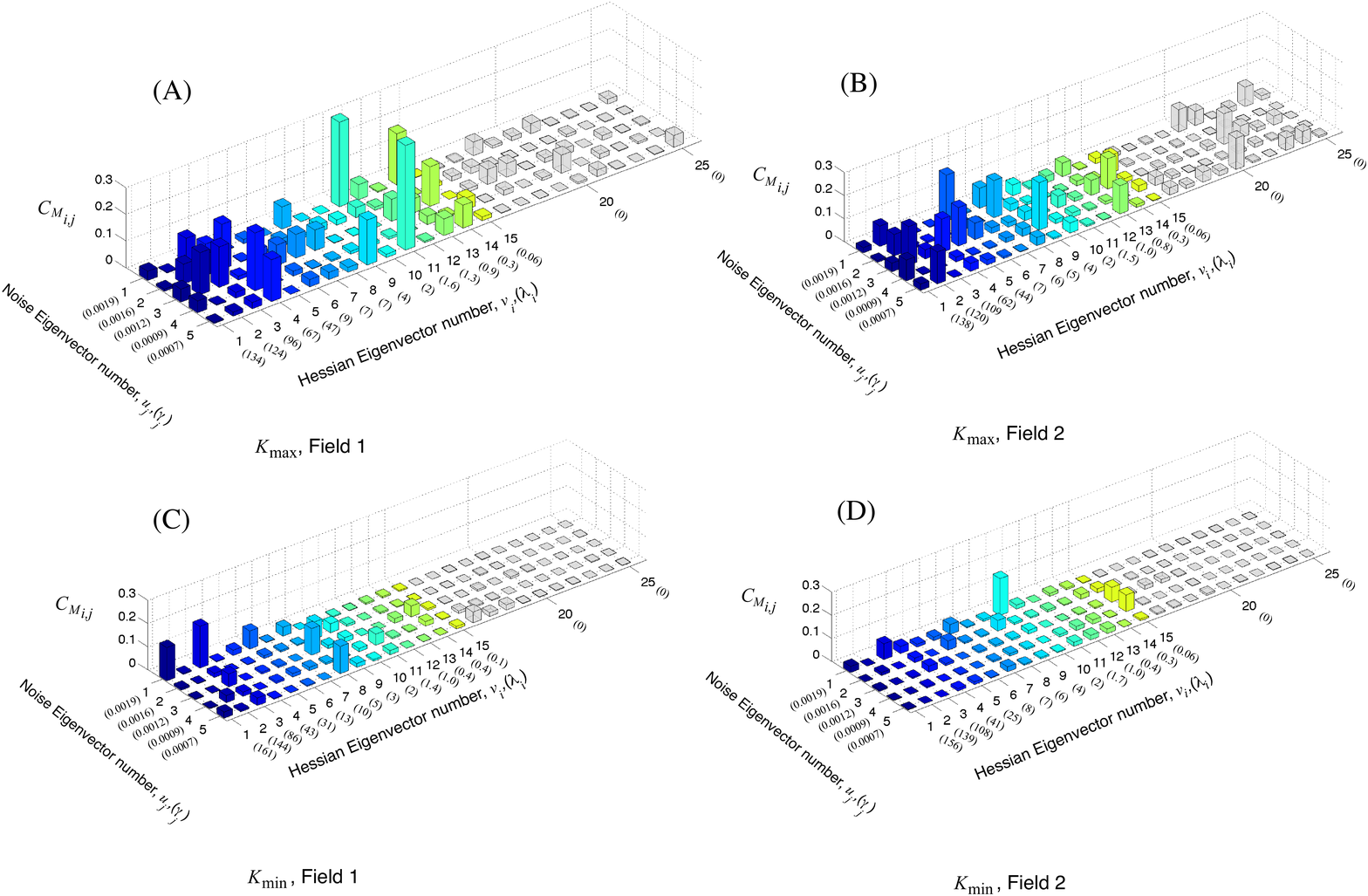}}
\caption{\label{fig:colored_cnot_mult} (color online) Multiplicative control noise and Hessian overlap terms  $C_{M_{i,j}}$ for mid-frequency noise in the CNOT gate.  The noise correlation time for the noise was $\alpha=0.1$. Overlap terms for the individual control field contributions are shown separately for worst robustness (Figs. (A) and (B)) and best robustness (Figs. (C) and (D)). Listed in parentheses are the associated eigenvalues of each eigenfunction. The eigenfunctions composing the Hessian nullspace are colored in gray. Robustness is achieved by diminishing the Hessian non-nullspace overlap with the noise spectrum, regardless of the overlap with the Hessian nullspace.}
\end{center}
\end{figure*}

Table \ref{tab:midfreq_results} presents the values of the robustness measure and fluence for three separate controls that are representative of robust ($K_{\mathrm{min}}$), average ($\langle K \rangle$), and non-robust ($K_{\mathrm{max}}$) controls in the presence of noise with a correlation time $\alpha = 0.1$. The fluence of each control field is a measure of its energy,
\begin{align}
\label{eq:fluence}
f_i &= \int_0^T \epsilon_i^2(t) dt, \quad i = 1,2.
\end{align}
For the case of the Hadamard gate there is a single field with fluence, $f$. The table shows an intuitive trend that higher fluence controls are less robust toward multiplicative noise, but interestingly the behavior is essentially the same for additive noise. Additionally, the overlap terms $C_{A_{i, j}}$ from Eq. (\ref{eq:K_eigendecomp_general1})  for robust and non-robust optimal fields were examined for performing the Hadamard gate for additive noise (Fig. \ref{fig:colored_add}), as well as $C_{M_{i, j}}$ from Eq. (\ref{eq:K_eigendecomp_general2}) with multiplicative noise  for CNOT gate  (Fig. \ref{fig:colored_cnot_mult}).  Figure \ref{fig:colored_add} illustrates the simple circumstances when robustness to noise is achieved by shifting the dynamics such that the noise spectrum contribution mainly lies in the Hessian's nullspace. In contrast, Figure \ref{fig:colored_cnot_mult} shows that robustness to multiplicative noise in each field performing the CNOT gate is achieved by reducing the partial overlap of the noise in the Hessian non-nullspace, regardless of the overlap in the Hessian nullspace.  This behavior is further explained by the rapid decrease of the Hessian eigenvalues (in parentheses), such that components beyond index $i\approx 6$ have little overall contribution to $K$. Such contrasting behavior between Figs. \ref{fig:colored_add} and \ref{fig:colored_cnot_mult} illustrates the variety of different ways in which a control can be robust to noise.

\begin{table}[htb]
 \caption{\label{tab:midfreq_results} Robustness and fluence for mid-frequency control noise, $\alpha = 0.1$ }
\begin{center}
\begin{tabular}{p{1.5cm} l l l c l l l }
 \hline
 \hline
&\multicolumn{7}{ c}{$W_{\mathrm{H}}$}\\

&\multicolumn{3}{c }{Add. Noise} & &\multicolumn{3}{c }{Mult. Noise} \\
& $K_{\mathrm{min}}$&  $\langle K \rangle $&  $K_{\mathrm{max}}$& & $K_{\mathrm{min}}$&  $\langle K \rangle $&  $K_{\mathrm{max}}$ \\
\cline{2-4}\cline{6-8}
 $K_{A/M}$ ($10^{-6}$) & $1.12 $& $1.17$  & $1.39$ & &$23.5 $  & $37.0 $ & $96.5 $\\[2pt]
  $f$ & $10.03 $& $17.39 $  & $42.94 $  & & $9.99 $  & $16.74 $ & $42.94 $\\[2pt]
\hline
&\multicolumn{7}{c}{$W_{\mathrm{CNOT}}$}\\
&\multicolumn{3}{c }{Add. Noise} & &\multicolumn{3}{c }{Mult. Noise} \\
& $K_{\mathrm{min}}$&  $\langle K \rangle $&  $K_{\mathrm{max}}$& &$K_{\mathrm{min}}$&  $\langle K \rangle $&  $K_{\mathrm{max}}$ \\
\cline{2-4}\cline{6-8}
$K_{A/M} $ ($10^{-6}$) & $17.6$& $17.7 $  & $17.9 $ & & $19.4 $  & $35.8 $ & $65.6$\\[2pt] 
  $f_1$ & $3.98$& $8.04$  & $19.18$ && $3.93$  & $5.82$ & $15.36$\\[2pt]
   $f_2$  & $ 4.97$& $11.04$  & $19.17$ && $4.34$  & $11.14$ & $15.03$\\[2pt]
\hline
\hline

\end{tabular}
\end{center}
\end{table}

\subsection{Low-frequency noise}
\label{sec:colored_noise}
Low-frequency noise ($\alpha \gg 1$) has a correlation time that is long compared to the timescale of the dynamics, and  can be treated as constant in time as $R(t,t') \approx R$. The mismatch in timescales for low-frequency noise has been exploited with many pulse-sequencing control techniques  \cite{viola_prl_1999, viola_pra_1998, Batista_annrevphyschem_2009, Viola_prl_2009, Viola_pra_2009}. This circumstance in the additive noise case leads to  
\begin{align}
\label{eq:KA_simp1}
K_{A} &=  \frac{R}{2} \sum_{i=1}^{N^2-1} \lambda_{i} V_{A_i},
\end{align}
where
\begin{align}
\label{eq:vi}
V_{A_i} &= \left( \int_0^T v_i(t) dt \right ) ^2.
\end{align}
The Hessian eigenfunctions with non-zero eigenvalues are typically highly oscillatory functions reflecting the system's dynamical sensitivity to the field, implying that time averaging over these eigenfunctions would lead to good robustness in this regime, as found in Figs. \ref{fig:averages}(A) and \ref{fig:averages}(B).

Similarly, for multiplicative noise, we have
\begin{align}
\label{eq:KM_simp1}
K_{M} &=   \frac{R_z}{2} \sum_{i=1}^{N^2-1} \lambda_{i} V_{M_i},
\end{align}
where
\begin{align}
\label{eq:Vim}
V_{M_i} &= \left(\int_0^T \epsilon(t)v_i(t)dt \right)^2.
\end{align}
The Hessian eigenfunctions with non-zero eigenvalues naturally reflect the key control field structure. Thus, the overlap in Eq. (\ref{eq:Vim}) is expected to be significant, which is reflected in the strong impact of multiplicative noise, over that of additive noise in Table \ref{tab:lowfreq} and in the ensembles in Fig. \ref{fig:averages}.

\begin{table}[htb]
 \caption{\label{tab:lowfreq}  Robustness and fluence for low-frequency control noise, $\alpha = 100$.}
\begin{center}
\begin{tabular}{p{1.5cm} l l l c l l l }
 \hline
 \hline
&\multicolumn{7}{ c}{$W = \mathrm{H}$}\\

&\multicolumn{3}{c }{Add. Noise} & &\multicolumn{3}{c }{Mult. Noise} \\
& $K_{\mathrm{min}}$&  $\langle K \rangle$ &  $K_{\mathrm{max}}$ & & $K_{\mathrm{min}}$&  $\langle K \rangle$ &  $K_{\mathrm{max}}$ \\
\cline{2-4}\cline{6-8}
 $K_{A/M}$ ($10^{-6}$) & $0.03 $& $0.30$  & $1.23$ && $74 $  & $128 $ & $333 $\\[2pt] 
  $f$ & $24.8 $& $30.03 $  & $20.86 $  &&  $17.0 $  & $17.3 $ & $35.3 $\\[2pt]
\hline
&\multicolumn{7}{c}{$W = \mathrm{CNOT}$}\\
&\multicolumn{3}{c }{Add. Noise} & &\multicolumn{3}{c }{Mult. Noise} \\
& $K_{min}$&  $\langle K \rangle $&  $K_{max}$& & $K_{min}$&  $\langle K \rangle $&  $K_{max}$ \\
\cline{2-4}\cline{6-8}
$K_{A/M} $ ($10^{-6}$) & $0.17$& $0.26 $  & $0.65 $ && $113 $  & $193 $ & $466$\\[2pt] 
  $f_1$ & $5.21$& $8.16$  & $13.68$ && $9.28$  & $8.13$ & $7.81$\\[2pt]
$f_2$ & $10.47$& $7.81$  & $14.85$ && $11.02$  & $6.08$ & $14.20$\\[2pt]
\hline
\hline
\end{tabular}\\
\end{center}
\end{table}

\subsection{White Noise}
Another limiting case for robustness occurs for Gaussian white noise (i.e., $\delta$-correlated), in which the power density spectrum covers the entire frequency domain. This case has been previously examined \cite{brif_robustness_2012, moore_pra_2012, Kosut_sco_2013}, 
and also the robustness scaling with respect to the system dimension has been studied for a class of variable-size systems with a particular dipole moment structure \cite{Kallush.njphys.1308.6128.2013}.  We briefly summarize the circumstances to demonstrate the contrast between the robustness behavior of different spectral regimes of noise . The robustness measure for additive white noise becomes
\begin{align}
\label{eq:white_add_noise}
K_{A} &= \frac{A^2}{2} \int_0^T \int_0^T \mathcal{H}_{\epsilon}(t,t') \delta(t-t') dt' dt  \nonumber \\
&= \frac{A^2}{2}  \text{Tr}[\mathcal{H}_{\epsilon}]  \nonumber \\
&= \frac{A^2 T}{4N}  \| \mu  \|^2, 
\end{align}
The fixed trace shows invariance to pulse shaping, and robustness can only be increased through a shorter operation time, $T$.

 Similarly, for multiplicative control noise the fixed Hessian trace leads to
\begin{align}
\label{eq:white_mult_noise}
K_{\epsilon, M} &= \frac{A^2}{2} \int_0^T \int_0^T \mathcal{H}_{\epsilon}(t,t') \epsilon(t)\epsilon(t') \delta(t-t') dt' dt \nonumber \\
&= \frac{A^2}{4N} \left \| \mu \right \|^2f,
\end{align}
in which case robustness can only be enhanced by decreasing the fluence $f$. In both cases, white noise offers little opportunity to enhance robustness.

\section{Conclusion}
\label{sec:conclusion}
This work utilized the control landscape Hessian to provide a general framework for quantifying the robustness of targeted unitary gate operations in the presence of random noise. Ensembles of randomly generated, fidelity-optimized controls revealed that distinct spectral regimes of noise exist where robustness quality is highly diverse. Numerical examination of low-frequency and mid-frequency control noise demonstrated that even though the \emph{total} landscape curvature around any optimal control point is fixed (i.e., the Hessian trace is invariant to the control for a given $T$), robust controls can still correspond to landscape domains possessing curvature that is favorable, with Hessian eigenfunctions oriented away from the disturbances due to noise.

The challenges faced upon seeking optimal robust controls are evident in Fig. \ref{fig:averages}, where the mean performance is $\langle J \rangle = J_0 + \langle K \rangle$, with $J_0 <  10^{-6}$ in the present work. Importantly, in the regime of weak noise, $\langle K \rangle$ scales as $A^2$ from the noise correlation function strength in  Eq. (\ref{eq:acf_forms}), with $A = 0.01$ chosen to represent $\sim 0.1\%$ of the optimal field amplitude. Based on these randomly sampled tests, robust performance requires that the value of $A$ should be further reduced, in particular for the CNOT gate, to ensure fault-tolerant operation. In addition, the left standard deviation $\sigma_l$ in Eq. (\ref{eq:std_l}) also scales as $A^2$, so a reduction in $A$ also leaves less room for optimal field enhancement of robustness. These insights into the robustness of controls are relevant to optimal control experiments, and a full assessment of this matter calls for further work exploring for optimally robust controls, as well as potential tradeoffs between fidelity and robustness. Finally, the landscape perspective draws attention to the equally important roles of control noise and system dynamics when considering robustness. Thus, for designed quantum devices (e.g., gates), balanced attention should be given to alternative system realizations and the associated control noise characteristics. 

\begin{acknowledgements}
This material is based upon work supported by the National Science Foundation Graduate Research Fellowship Program under Grant No. (DGE 1148900), National Science Foundation (CHE-1058644) and ARO-MURI (W911NF-11-1-2068). This work is also supported by the Laboratory Directed Research and Development program at Sandia National Laboratories. Sandia is a multi-program laboratory managed and operated by Sandia Corporation, a wholly owned subsidiary of Lockheed Martin Corporation, for the United States Department of Energy's National Nuclear Security Administration under contract DE-AC04-94AL85000. RBW acknowledges support from the NSFC (Grant No. 61374091 and 61134008).
\end{acknowledgements}

\bibliographystyle{apsrev4-1}
\bibliography{hocker_revised}	

\end{document}